\documentstyle[aps,draft,epsf]{revtex}
\begin{document}
\draft
\title{Semiclassical Quantization for the Spherically Symmetric Systems under an
Aharonov-Bohm magnetic flux}
\author{W.F. Kao\thanks{%
wfgore@cc.nctu.edu.tw}, P.G. Luan, and D.H. Lin \thanks{
e-mail: dhlin@cc.nctu.edu.tw}}
\address{Institute of Physics,\\
Chiao Tung University, Hsin Chu, 30043, Taiwan}
\date{\today}
\maketitle

\begin{abstract}
The semiclassical quantization rule is derived for a system with a
spherically symmetric potential $V(r) \sim r^{\nu }$ $(-2<\nu
<\infty )$ and an Aharonov-Bohm magnetic flux. Numerical results
are presented and compared with known results for models with $\nu
= -1,0,2,\infty$. It is shown that the results provided by our
method are in good agreement with previous results. One expects
that the semiclassical quantization rule shown in this paper will
provide a good approximation for all principle quantum number even
the rule is derived in the large principal quantum number limit $n
\gg 1$. We also discuss the power parameter $\nu $ dependence of
the energy spectra pattern in this paper.
\end{abstract}
%\maketitle
%\newpage \tolerance=10000

%\begin{document}
%\pagestyle{empty}                                      %%%To be commented
%\preprint{ \font\fortssbx=cmssbx10 scaled \magstep2 \hbox to
%\hsize{ \hfill$\raise .5cm\vtop{
%                \hbox{NCTU-HEP-0103}}$}}
%\draft
%
%\vfill

%\title{On Analytic Properties of the Photon Polarization Function and pair production width in a
%Background Magnetic Field  }

%\author{W. F. Kao and Guey-Lin Lin} %

%%\address{Institute of Physics, National Chiao Tung University,
%%Hsinchu 300, Taiwan}
%\affiliation{Institute of Physics, Chiao Tung University, Hsin Chu, Taiwan}
%\date{\today}
%
%\vskip -1cm
%
%\vfill
%
%%\maketitle
%\begin{abstract}

%\end{abstract}
%
%\vfill
%
%\pacs{PACS numbers: 12.20.Ds, 11.55.Fv}
%\maketitle
%
%\narrowtext
%
%\pagestyle{plain}

%\documentstyle[eqsecnum,aps]{revtex}
%%%%%%%%%%%%%%%%%%%%%%%%%%%%%%%%%%%%%%%%%%%%%%%%%%%%%%%%%%%%%%%%%%%%%%%%%%%%%%%%%%%%%%%%%%%%%%%%%%%%%%%%%%%%%%%%%%%%%%%%%%%%
%TCIDATA{OutputFilter=Latex.dll} %TCIDATA{LastRevised=Fri Nov 09 11:41:492001} %TCIDATA{<DEFANGED_META NAME="GraphicsSave" CONTENT="32">}
%TCIDATA{Language=American English} %TCIDATA{CSTFile=revtex.cst}

%\pagestyle{plain}

\section{Introduction}

In the past 20 years, Aharonov-Bohm (A-B) effect, a topological
non-local physical effect at the quantum level, has been of much
interest in the studies of cosmic string \cite{1}, $(2+1)$-$D$
gravity theories \cite{2} and especially in the context of anyon
\cite{3}, which has shed light on the understanding of the
phenomenon of the fractional quantum Hall effect \cite {4,5,6,7},
superconductivity \cite{7,8}, repulsive Bose gases \cite{9}, and
so forth. There are only a few models coupled to different
potentials along with an A-B magnetic flux that can be solved
exactly. For the system with both an A-B magnetic flux and a
spherically symmetric potential of the form $ V(r)=\lambda r^{\nu
}(-2<\nu <\infty )$, the solvable models known to us
include the cases with the parameter $\nu =-1,0,2,\infty $ \cite{10,11,12,13}.%
Here $\lambda$ is a constant parameter. Note that when $\nu =-1$,
it is a system with both an A-B magnetic flux and a Coulomb
potential (A-B-C) \cite {10,11,12}. This system describes the
relative motion of two charged
particles, with one of them carrying electric charge and magnetic flux $%
(-q,-\Phi /Z)$ while the other one carrying $(Zq,\Phi )$. Here $Z
( \neq 0)$ is a non-vanishing real number. This system is of much
interest in many different areas \cite{12}.

In the past three decades, much progress has been made in the
semiclassical methods toward the understanding of these systems.
These kinds of semiclassical methods provide with us a powerful
approximation tool in different areas in order to extract useful
information from various unsolved problems including the
quantization of the classical chaotic systems \cite{Gu}, deformed
atomic nuclei, asymmetric fission nuclei \cite{St}, semiclassical
quantum dots, and weak localization in mesoscopic systems
\cite{Ma}. In this paper, we will consider a generalized system
with both an A-B magnetic flux and a spherically symmetric
potential of the form mentioned above. The set of the parameters
($\lambda$, $\nu$)
will be discussed in the following ranges (i) $(\lambda <0,\;-2<\nu <0)$ and (ii) $%
(\lambda >0,\;\nu >0)$. We will derive a semiclassical
quantization rule of the approximated energy spectra for this set
of parameters. The distribution tendency of the energy spectra on
different values of the parameter $\nu $ will also be
given. By comparing with the known results, including the models with $%
\nu=-1,0,2$, we find that our method agrees with these exact
results. In addition, for the exactly solvable model with
$\nu=\infty$, the difference between the exact and semiclassical
results will be shown to be very small from a numerical
computation. Therefore, we are confident in that our formulae will
also provide a good approximation for the two ranges of parameters
mentioned above where $\nu\neq -1,0,2,\infty$.

This paper is organized as follows. In section II, we will derive
the semiclassical quantization rule of the A-B effect under a
spherically symmetric potential. In particular, we will first
derive the non-integrable phase factor of the Green's function due
to the A-B effect in a spherically symmetric system. The
corresponding radial Schr\"{o}dinger equation will also be derived
accordingly. The semiclassical wave functions will also be derived
according to the semiclassical consideration of the Bohr's
corresponding principle. Consequently, the quantization rule can
thus be obtained by comparing with the well-known WKB phase. We
will also study the distribution dependence of energy spectra in
various models in section III. The effect of magnetic flux will
also be discussed and emphasized in this section. Finally, in
section IV, some conclusions will be drawn. In order to provide a
self-contained information, we will show the WKB matching
condition of the semiclassical wave functions in the appendix.

\section{Semiclassical Quantization Rule of the A-B effect with a
Spherically Symmetric Potential}

The fixed-energy Green's function $G^{0}({\bf r,r}^{\prime };E)$
for a charged particle with mass $m$ propagating from ${\bf r}$ to ${\bf r}%
^{\prime }$ satisfies the Schr\"{o}dinger equation
\begin{equation}
\left[ E-H_{0}({\bf r},\frac{\hbar }{i}{\bf \nabla )}\right] G^{0}({\bf r,r}%
^{\prime };E)=\delta ^{3}({\bf r-r}^{\prime }),  \label{a1}
\end{equation}
where the system Hamiltonian is given by $H_{0}=-$ $\hbar ^{2}{\bf \nabla }%
^{2}/2m+V(r)$ as usual. In the spherically symmetric cases, the angular
decomposition of the Green's function can be written as
\begin{equation}
G^{0}({\bf r,r}^{\prime };E)=\sum_{l=0}^{\infty }\sum_{k=-l}^{l}G_{l}^{0}(r%
{\bf ,}r^{\prime };E)Y_{lk}(\theta ,\varphi )Y_{lk}^{\ast }(\theta ^{\prime
},\varphi ^{\prime })  \label{a2}
\end{equation}
with $Y_{lk}$ the well-known spherical harmonics. As a result, the left hand
side of Eq. (\ref{a1}) can be brought to the following form
\[
\left\{ E-\sum_{l=0}^{\infty }\sum_{k=-l}^{l}\left[ -\frac{\hbar ^{2}}{2m}%
\left( \frac{d^{2}}{dr^{2}}+\frac{2}{r}\frac{d}{dr}\right) +\frac{%
l(l+1)\hbar ^{2}}{2mr^{2}}\right] -V(r)\right\}
\]
\begin{equation}
\times G_{l}^{0}(r{\bf ,}r^{\prime };E)Y_{lk}(\theta ,\varphi )Y_{lk}^{\ast
}(\theta ^{\prime },\varphi ^{\prime }).  \label{a3}
\end{equation}
For a charged particle in a magnetic field, the Green's function
$G$ is related to $G^{0}$ by the following equation
\begin{equation}
G({\bf r,r}^{\prime };E)=G^{0}({\bf r,r}^{\prime };E)e^{\frac{ie}{\hbar c}%
\int_{{\bf r}^{\prime }}^{{\bf r}}{\bf A}({\bf \tilde{r})\cdot }d{\bf \tilde{%
r}}},  \label{a4}
\end{equation}
with a globally path-dependent non-integrable phase factor
\cite{14,15} given above. Here we have used the vector potential
${\bf A}({\bf \tilde{r})} $ to represent the magnetic field. For
the Aharonov-Bohm magnetic flux under consideration, the vector
potential can be written as
\begin{equation}
{\bf A(x)=}\left\{
\begin{array}{l}
\frac{1}{2}B\rho \hat{e}_{\varphi }\qquad \qquad (\rho <\epsilon ) \\
\frac{1}{2}B\frac{\epsilon ^{2}}{\rho }\hat{e}_{\varphi }=\frac{\Phi }{2\pi
\rho }\hat{e}_{\varphi }\quad (\rho >\epsilon )
\end{array}
\right. ,  \label{a5}
\end{equation}
where the two-dimensional radial length is defined as $\rho ^{2}=x^{2}+y^{2}$
as usual. Moreover, $\hat{e}_{\varphi }$ is the unit vector of coordinate $%
\varphi $, $\epsilon $ is the radius of region where magnetic
field exists. Hence the total magnetic flux is given by $\Phi =\pi
\epsilon ^{2}B$. Note that the associated magnetic field lines are
confined inside a tube, with radius $\epsilon $, along the
$z$-axis. Along the region without magnetic field, the
path-dependent non-integrable phase factor is given by
\begin{equation}
e^{-i\mu _{0}\int_{P}^{\lambda }d\lambda ^{\prime }\dot{\varphi}(\lambda
^{\prime })},  \label{a7}
\end{equation}
where we have used the subscript $P$ to represent the path
dependent nature of phase factor and we have denoted
$\dot{\varphi}(\lambda ^{\prime })=d\varphi /d\lambda ^{\prime }$.
Also, $\mu _{0}=-2eg/\hbar c$ is a dimensionless number defined by
$\Phi =4\pi g$. The minus sign is a matter of convention.
According to the discussion in Ref. \cite{15}, only phase factors
with closed-loop contour are considered where the description of
electromagnetic phenomenon are complete. Hence, we have
\begin{equation}
n=\frac{1}{2\pi }\int_{P}^{\lambda }d\lambda ^{\prime }\dot{\varphi}(\lambda
^{\prime })  \label{a8}
\end{equation}
with integer values $n$ corresponding the winding number. The
magnetic interaction is therefore purely topological. Therefore
the nonintegrable phase factor becomes
\begin{equation}
e^{-i\mu _{0}\left( 2n\pi \right) }.  \label{a9}
\end{equation}
With the help of the equality between the associated Legendre polynomial $%
P_{\nu }^{\mu }(z)$ and the Jacobi function $P_{n}^{\left( \alpha ,\beta
\right) }(z)$ \cite{16,17}, we find that
\begin{equation}
P_{l}^{k}(\cos \theta )=(-1)^{k}\frac{\Gamma (l+k+1)}{\Gamma (l+1)}\left(
\cos \frac{\theta }{2}\sin \frac{\theta }{2}\right) ^{k}P_{l-k}^{\left(
k,k\right) }(\cos \theta ).  \label{a10}
\end{equation}
Therefore the angular part of the Green's function in the
expression (\ref{a3}) can be turned into the following form
\[
\sum_{k{\bf =-}l}^{l}Y_{lk}(\theta ,\varphi )Y_{lk}^{\ast }(\theta ^{\prime
},\varphi ^{\prime })=\sum_{k{\bf =-}l}^{l}\frac{2l+1}{4\pi }\frac{\Gamma
\left( l-k+1\right) }{\Gamma \left( l+k+1\right) }P_{l}^{k}(\cos \theta
)P_{l}^{k}(\cos \theta ^{\prime })e^{ik(\varphi -\varphi ^{\prime })}
\]
\[
=\sum_{k{\bf =-}l}^{l}\left[ \frac{2l+1}{4\pi }\frac{\Gamma \left(
l-k+1\right) \Gamma \left( l+k+1\right) }{\Gamma ^{2}\left( l+1\right) }%
\right] \left( \cos \frac{\theta }{2}\cos \frac{\theta ^{\prime }}{2}\sin
\frac{\theta }{2}\sin \frac{\theta ^{\prime }}{2}\right) ^{k}
\]
\begin{equation}
\times P_{l-k}^{\left( k,k\right) }(\cos \theta )P_{l-k}^{\left( k,k\right)
}(\cos \theta ^{\prime })e^{ik\left( \varphi -\varphi ^{\prime }\right) }.
\label{a11}
\end{equation}
In order to include the non-integrable phase factor due to the A-B
effect, we will change the index $l$ into $q$ related by the
definition $l-k=q$. As a result one can rewrite the Eq. (\ref{a3})
as
\[
\left\{ E-\sum_{q=0}^{\infty }\sum_{k=-\infty }^{\infty }\left[ -\frac{\hbar
^{2}}{2m}\left( \frac{d^{2}}{dr^{2}}+\frac{2}{r}\frac{d}{dr}\right) +\frac{%
(q+k)(q+k+1)\hbar ^{2}}{2mr^{2}}\right] -V(r)\right\}
\]
\[
\times G_{q+k}^{0}(r{\bf ,}r^{\prime };E)\left[ \frac{2(q+k)+1}{4\pi }\frac{%
\Gamma \left( q+1\right) \Gamma \left( q+2k+1\right) }{\Gamma ^{2}\left(
q+k+1\right) }\right] \left( \cos \frac{\theta }{2}\cos \frac{\theta
^{\prime }}{2}\sin \frac{\theta }{2}\sin \frac{\theta ^{\prime }}{2}\right)
^{k}
\]
\begin{equation}
\times P_{q}^{\left( k,k\right) }(\cos \theta )P_{q}^{\left( k,k\right)
}(\cos \theta ^{\prime })e^{ik\left( \varphi -\varphi ^{\prime }\right) }.
\label{a12}
\end{equation}
In addition, the non-integrable phase in Eq. (\ref{a9}) can now be
included with the help of the Poisson's summation formula (p.124,
\cite{17a})
\begin{equation}
\sum_{k=-\infty }^{\infty }f(k)=\int_{-\infty }^{\infty }dy\sum_{n=-\infty
}^{\infty }e^{2\pi nyi}f(y).  \label{a13}
\end{equation}
Therefore, the expression (\ref{a12}) can be written as
\[
\left\{ E-\sum_{q=0}^{\infty }\int dz\sum_{k=-\infty }^{\infty }\left[ -%
\frac{\hbar ^{2}}{2m}\left( \frac{d^{2}}{dr^{2}}+\frac{2}{r}\frac{d}{dr}%
\right) +\frac{(q+z)(q+z+1)\hbar ^{2}}{2mr^{2}}\right] -V(r)\right\}
\]
\[
\times G_{q+z}(r{\bf ,}r^{\prime };E)\left[ \frac{2(q+z)+1}{4\pi }\frac{%
\Gamma \left( q+1\right) \Gamma \left( q+2z+1\right) }{\Gamma ^{2}\left(
q+z+1\right) }\right] \left( \cos \frac{\theta }{2}\cos \frac{\theta
^{\prime }}{2}\sin \frac{\theta }{2}\sin \frac{\theta ^{\prime }}{2}\right)
^{z}
\]
\begin{equation}
\times P_{q}^{\left( z,z\right) }(\cos \theta )P_{q}^{\left( z,z\right)
}(\cos \theta ^{\prime })e^{i(z-\mu _{0})\left( \varphi +2k\pi -\varphi
^{\prime }\right) },  \label{a14}
\end{equation}
where the superscript $^{0}$ in $G_{q+k}^{0}$ has been suppressed
to reflect the inclusion of the A-B effect. The summation over all
indices $k$ forces $z= \mu _{0}$ modulo an arbitrary integer
number. Therefore, one has
\[
\left\{ E-\sum_{q=0}^{\infty }\sum_{k=-\infty }^{\infty }\left[ -\frac{\hbar
^{2}}{2m}\left( \frac{d^{2}}{dr^{2}}+\frac{2}{r}\frac{d}{dr}\right) +\frac{%
(q+\left| k+\mu _{0}\right| )(q+\left| k+\mu _{0}\right| +1)\hbar ^{2}}{%
2mr^{2}}\right] -V(r)\right\}
\]
\[
\times G_{q+\left| k+\mu _{0}\right| }(r{\bf ,}r^{\prime };E)\left\{ \frac{%
\left[ 2\left( q+\left| k+\mu _{0}\right| \right) +1\right] }{4\pi }\frac{%
\Gamma \left( q+1\right) \Gamma \left( 2\left| k+\mu _{0}\right| +q+1\right)
}{\Gamma ^{2}\left( \left| k+\mu _{0}\right| +q+1\right) }\right\}
e^{ik\left( \varphi -\varphi ^{\prime }\right) }
\]
\begin{equation}
\times \left( \cos \theta /2\cos \theta ^{\prime }/2\sin \theta /2\sin
\theta ^{\prime }/2\right) ^{\left| k+\mu _{0}\right| }P_{q}^{\left( \left|
k+\mu _{0}\right| ,\left| k+\mu _{0}\right| \right) }(\cos \theta
)P_{q}^{\left( \left| k+\mu _{0}\right| ,\left| k+\mu _{0}\right| \right)
}(\cos \theta ^{\prime }).  \label{a15}
\end{equation}
Note that the influence of the A-B effect to the radial Green's
function is
to replace the integer quantum number $l$ with a fractional quantum number $%
q+\left| k+\mu _{0}\right| $. Analogously the same procedure can
be applied to the delta function $\delta ^{3}({\bf r-r}^{\prime
})$ in the \ rhs of the Eq. (\ref{a1}) with the help of the
following solid angle representation of the $\delta $ function
\begin{equation}
\delta \left( \Omega -\Omega ^{\prime }\right) =\sum_{l=0}^{\infty
}\sum_{k=-l}^{l}Y_{lk}(\theta ,\varphi )Y_{lk}^{\ast }(\theta ^{\prime
},\varphi ^{\prime }).  \label{a161}
\end{equation}
Therefore, for the set of the fixed quantum numbers $(q,k)$ one
can show that the radial Green's function satisfies
\[
\left\{ E-\left[ -\frac{\hbar ^{2}}{2m}\left( \frac{d^{2}}{dr^{2}}+\frac{2}{r%
}\frac{d}{dr}\right) +\frac{(q+\left| k+\mu _{0}\right| )(q+\left| k+\mu
_{0}\right| +1)\hbar ^{2}}{2mr^{2}}\right] -V(r)\right\}
\]
\begin{equation}
\times G_{q+\left| k+\mu _{0}\right| }(r{\bf ,}r^{\prime };E)=\delta (r{\bf -%
}r^{\prime }).  \label{a16}
\end{equation}
As a result, the corresponding radial wave equation reads
\begin{equation}
\frac{\hbar ^{2}}{2m}\frac{d^{2}}{dr^{2}}u_{\gamma }(r)+\left[ E-\left( V(r)+%
\frac{\hbar ^{2}}{2m}\frac{\gamma (\gamma +1)}{r^{2}}\right) \right]
u_{\gamma }(r)=0,  \label{a17}
\end{equation}
where we have set $\gamma =q+\left| k+\mu _{0}\right| $, and $u_{\gamma
}(r)\equiv rR_{\tilde{n}\gamma }(r)$. Obviously, $R_{\tilde{n}\gamma }$
satisfies the spherical Bessel equation
\begin{equation}
\left[ \frac{d^{2}}{dr^{2}}+\frac{2}{r}\frac{d}{dr}+\left( \kappa ^{2}-U(r)-%
\frac{\gamma (\gamma +1)}{r^{2}}\right) \right] R_{\tilde{n}\gamma }(r)=0
\label{a18}
\end{equation}
with the definitions $\kappa =\sqrt{2mE/\hbar ^{2}}$ and the reduced
potential $U(r)=2mV(r)/\hbar ^{2}$. For simplicity, we have written $R_{%
\tilde{n}\gamma }(r)$ instead of $R_{\tilde{n},q,k}(r)$ in which each set $(%
\tilde{n},q,k)$ denotes a quantum state. Hence the A-B effect
reflects itself by the coupling to the angular momentum in radial
Green's function which turns the integer quantum number into a
fractional one.

To find the semiclassical quantization rule, let us first consider
the asymptotic form of the bound state wave functions of a charge
article moving in a spherically symmetric potential of the form
$V(r)=\lambda r^{\nu }(\lambda <0,-2<\nu <0)$\ under an A-B
magnetic flux for the energy limit $E\rightarrow 0$. Due to the
Bohr corresponding principle, this stands for the semiclassical
approximation since there are infinitely densed energy levels near
$E\rightarrow 0^{-}$. According to the Eq. (\ref{a17}), the
asymptotic wave equation reads, in the $E \rightarrow 0^-$ limit,
\begin{equation}
\frac{\hbar ^{2}}{2m}\frac{d^{2}}{dr^{2}}u_{\gamma }(r)-\left( \lambda
r^{\nu }+\frac{\hbar ^{2}}{2m}\frac{\gamma (\gamma +1)}{r^{2}}\right)
u_{\gamma }(r)=0.  \label{a19}
\end{equation}
We can also perform the following transformations:
\begin{equation}
\rho =r\left( \frac{2m\left| \lambda \right| }{\hbar ^{2}}\right) ^{1/(\nu
+2)},\;u(r)=W(\rho ).  \label{a20}
\end{equation}
Consequently, the Eq. (\ref{a19}) yields
\begin{equation}
\frac{d^{2}W}{d\rho ^{2}}+\left[ \rho ^{\nu }-\frac{\gamma (\gamma +1)}{\rho
^{2}}\right] W=0,  \label{a21}
\end{equation}
which can be further reduced with the help of the following change of
variables
\begin{equation}
z=\frac{2}{\nu +2}\rho ^{(\nu +2)/2},\;W(\rho )=z^{1/(\nu +2)}v(z).
\label{a22}
\end{equation}
As a result, the Eq. (\ref{a21}) becomes
\begin{equation}
\frac{d^{2}v}{dz^{2}}+\frac{1}{z}\frac{dv}{dz}+\left[ 1-\left( \frac{2\gamma
+1}{\nu +2}\right) ^{2}\frac{1}{z^{2}}\right] v=0.  \label{a23}
\end{equation}
This is exactly the Bessel's equation of integral order $\nu_1
\equiv \left( 2\gamma +1\right) /\left( \nu +2\right) $. The
boundary condition (B.C.) of the function $ u(r)$ in the Eq.
(\ref{a17}) is simply $u(0)=0$. Therefore, the corresponding B.C.
of $v(z)$ in the Eq. (\ref{a23}) is $v(0)=0$. The Bessel function
of the first kind is known to be the solution of the Bessel's
equation. Therefore, by imposing the B.C. appropriately, one can
show that
\begin{equation}
v(z)=J_{\nu_1} (z)   \label{a24}
\end{equation}
is the solution of the Eq. (\ref{a23}) with the prescribed boundary
condition. Therefore, the solution of the radial wave equation near $E
\rightarrow 0$ becomes
\begin{equation}
u(r)=W(\rho )=z^{1/(\nu +2)}J_{\nu_1} (z).   \label{a25}
\end{equation}
>From the asymptotic behavior of the Bessel function near
$r\rightarrow 0$, or equivalently $\rho \rightarrow 0$ and
$z\rightarrow 0$, one can show that
\begin{equation}
u(r)\sim z^{1/(\nu +2)}z^{\left( 2\gamma +1\right) /\left( \nu +2\right)
}\sim z^{\left( 2\gamma +2\right) /\left( \nu +2\right) }\sim r^{\gamma +1}.
\label{a26}
\end{equation}
On the other hand, from the asymptotic behavior of the Bessel function
approaching $r\rightarrow \infty ,$
\begin{equation}
J_{\alpha }(z)\rightarrow \sqrt{\frac{2}{\pi z}}\cos \left( z-\frac{\alpha
\pi }{2}-\frac{\pi }{4}\right) ,  \label{a27}
\end{equation}
one can show that
\[
u(r)\sim z^{1/(\nu +2)}\sqrt{\frac{2}{\pi z}}\cos \left( z-\nu_1 \frac{\pi }{%
2}-\frac{\pi }{4}\right)
\]
\begin{equation}
\sim \rho ^{-\nu /4}\cos \left( \frac{2}{\nu +2}\rho ^{\left( \nu +2\right)
/2}-\nu_1 \frac{\pi }{2}-\frac{\pi }{4}\right) .  \label{a28}
\end{equation}
Note that one can also compute the following integral, near the
limit $E\rightarrow 0$, and show that the following identities
hold:
\[
\int_{0}^{r}\sqrt{\frac{2m}{\hbar ^{2}}(E-V(r))}dr=\left( \frac{2m\left|
\lambda \right| }{\hbar ^{2}}\right) ^{1/2}\int_{0}^{r}r^{\nu /2}dr
\]
\begin{equation}
=\left( \frac{2m\left| \lambda \right| }{\hbar ^{2}}\right) ^{1/2}\frac{2}{
\nu +2}r^{\left( \nu +2\right) /2}=\frac{2}{\nu +2}\rho ^{\left( \nu
+2\right) /2}.  \label{a29}
\end{equation}
It follows that, in the limit $E\rightarrow 0$ and $r\rightarrow
\infty $,
\[
u(r)\sim r^{-\nu /4}\cos \left( \int_{0}^{r}\sqrt{\frac{2m}{\hbar ^{2}}
(E-V(r))}-\nu_1 \frac{\pi }{2}-\frac{\pi }{4}\right)
\]

\begin{equation}
\sim r^{-\nu /4}\sin \left( \int_{0}^{r}\sqrt{\frac{2m}{\hbar ^{2}}(E-V(r))}%
- \nu_1 \frac{\pi}{2} +\frac{\pi }{4}\right) ,  \label{a30}
\end{equation}
where $V(r)=\lambda r^{\nu }(\lambda <0,-2<\nu <0).$ If we take
the integration upper bound $r$ as the classical turning point
$r_{c}$ where $V(r_{c})=E$ , the phase of $u(r)$ can be shown to
be the WKB phase (see Appendix A for details)
\begin{equation}
u(r_{c})\propto  \sin \left[ \left( n+\frac{3}{4}\right) \pi
\right] . \label{a31}
\end{equation}
Consequently, from comparing the equations (\ref{a30}) and
(\ref{a31}), one can extract the following quantization condition
\begin{equation}
\int_{0}^{r_{c}}\sqrt{2m(E-V(r))}=\left[ n+\frac{2\gamma +\nu +3}{ 2(\nu +2)}
\right] \pi \hbar ,\;\; {\rm for} \; n=0,1,2,3\cdots .  \label{a32}
\end{equation}
Here $n$ is the radial quantum number. Although the Eq.
(\ref{a30}) is obtained in the limit $E\rightarrow 0$, or
equivalently in the large quantum number where $n\gg 1$, above
result can still be extended to all possible values of $n$. In
fact, the integral in the Eq. (\ref{a32}) can be written in terms
of analytic form. Indeed, with the help of the following change of
variables
\begin{equation}
\frac{\lambda }{E}r^{\nu }=\csc ^{2}\xi ,  \label{a33}
\end{equation}
one can re-write the following integral as
\begin{equation}
\int_{0}^{r_{c}}\sqrt{2m(E-V(r))}=-\frac{2}{\nu }\left( \frac{E}{\lambda }%
\right) ^{1/\nu }\sqrt{2m\left| E\right| }\int_{0}^{\pi /2}\cos ^{2}\xi
(\sin \xi )^{-2/\nu -2}d\xi .  \label{a34}
\end{equation}
In addition, with the help of the following formula (see, for example, Ref.
\cite{18}, p8),
\begin{equation}
\int_{0}^{\pi /2}\cos ^{2q-1}z\sin ^{2p-1}zdz=\frac{\Gamma (p)\Gamma (q)}{%
2\Gamma (p+q)},  \label{a35}
\end{equation}
one has
\begin{equation}
\int_{0}^{r_{c}}\sqrt{2m(E-V(r))}=-\frac{2}{\nu }\left( \frac{E}{\lambda }%
\right) ^{1/\nu }\sqrt{2m\left| E\right| }\frac{\sqrt{\pi }}{4}\frac{\Gamma
\left( -\frac{1}{\nu }-\frac{1}{2}\right) }{\Gamma \left( 1-\frac{1}{\nu }%
\right) }.  \label{a36}
\end{equation}
Inserting the result of the Eq. (\ref{a36}) into the Eq.
(\ref{a32}), one has
\[
E=-\left| \lambda \right| ^{2/(\nu +2)}\left( \frac{\hbar ^{2}}{2m}\right)
^{\nu /(\nu +2)}
\]
\begin{equation}
\cdot \left[ 2\left| \nu \right| \sqrt{\pi }\left( n+\frac{2\left( q+\left|
k+\mu _{0}\right| \right) +\nu +3}{2\nu +4}\right) \cdot \frac{\Gamma \left(
1-\frac{1}{\nu }\right) }{\Gamma \left( -\frac{1}{\nu }-\frac{1}{2}\right) } %
\right] ^{2\nu /(\nu +2)} ,  \label{a37}
\end{equation}
where the ranges of the parameters are $\lambda ,E<0,\;-2<\nu
<0,\;n,q=0,1,2,\cdots ,$ and $-\infty <k<\infty $. For example,
with the potential of the form $V(r)=-e^{2}/r,$ one has
\begin{equation}
E_{n,q,k}=-mc^{2}\frac{\alpha ^{2}}{2\left[ n+q+\left| k+\mu
_{0}\right| +1 \right] ^{2}} . \label{a38}
\end{equation}
Here $\alpha =e^{2}/\hbar c$ denotes the fine structure constant.
This agrees with the exact result given in the Ref. \cite{11}. We
see that the A-B effect has changed the splitting of energy levels
although the electron moves in the absence of the magnetic field.
In addition, when the flux is quantized, namely, $4\pi g=(2\pi
\hbar c/e) \times $ integer, $\left| k+\mu _{0}\right| $ is an
integer and hence the spectrum is the same as the energy spectrum
of the pure hydrogen atom.

To obtain the semiclassical quantization rule for all positive power $\nu >0$
of the potential $V(r)=\lambda r^{\nu }$, one can perform the following
change of variable
\begin{equation}
\rho =r^{\alpha },\;u(r)=\rho ^{\beta }v(\rho )  \label{a39}
\end{equation}
and show that the Eq. (\ref{a17}), becomes
\[
\frac{d^{2}u}{dr^{2}}
\]
\begin{equation}
=\alpha ^{2}\rho ^{2+\beta -2/\alpha }\frac{d^{2}v(\rho )}{d\rho ^{2}}%
+\alpha ^{2}(2\beta +1-\frac{1}{\alpha })\rho ^{1+\beta -2/\alpha }\frac{%
dv(\rho )}{d\rho }+\alpha ^{2}\beta (\beta -\frac{1}{\alpha })\rho ^{\beta
-2/\alpha }v(\rho ).  \label{a40}
\end{equation}
Note that the different ranges $\nu >0$ and $\nu <0$ can be
properly adjusted when the parameters $\alpha$ and  $\beta $ are
chosen appropriately \cite {11a}. In addition, if we set
\begin{equation}
\alpha =-\frac{\nu }{\nu ^{\prime }}, \quad \beta =-\frac{1}{2}\left( 1+\frac{\nu
^{\prime }}{\nu }\right) ,  \label{a41}
\end{equation}
the term $dv/d\rho $ in the Eq. (\ref{a40}) disappears. Inserting
this back into the Eq. (\ref{a39}) and then the Eq. (\ref{a17}),
one has
\[
\frac{\hbar ^{2}}{2m}\rho ^{2+\nu ^{\prime }+2\nu ^{\prime }/\nu }\frac{%
d^{2}v}{d\rho ^{2}}+\left[ -\lambda \left( \frac{\nu ^{\prime }}{\nu }%
\right) ^{2}+E\left( \frac{\nu ^{\prime }}{\nu }\right) ^{2}\rho ^{\nu
^{\prime }}\right] v
\]
\begin{equation}
-\frac{\hbar ^{2}}{2m}\rho ^{-2}\left[ \gamma (\gamma +1)\left( \frac{\nu
^{\prime }}{\nu }\right) ^{2}+\frac{1}{4}\left( \frac{\nu ^{\prime }}{\nu }%
\right) ^{2}-\frac{1}{4}\right] \rho ^{2+\nu ^{\prime }+2\nu ^{\prime }/\nu
}v=0.  \label{a42}
\end{equation}
If we choose $\nu ^{\prime }=-2\nu /(2+\nu )$, the above equation reduces to
\begin{equation}
\frac{\hbar ^{2}}{2m}\frac{d^{2}v}{d\rho ^{2}}+\left[ E^{\prime }-\lambda
^{\prime }\rho ^{\nu ^{\prime }}-\gamma ^{\prime }(\gamma ^{\prime }+1)\frac{%
\hbar ^{2}}{2m\rho ^{2}}\right] v=0  \label{a44}
\end{equation}
with the following relations linking different parameters
\begin{equation}
\left\{
\begin{array}{l}
\nu ^{\prime }=-\frac{2\nu }{(2+\nu )} \\
E^{\prime }=-\lambda \left( \frac{\nu ^{\prime }}{\nu }\right) ^{2} \\
\lambda ^{\prime }=-E\left( \frac{\nu ^{\prime }}{\nu }\right) ^{2} \\
\gamma ^{\prime }=-(\gamma +\frac{1}{2})\frac{\nu ^{\prime }}{\nu }-\frac{1}{%
2}=\frac{2\gamma +1}{\nu +2}-\frac{1}{2}
\end{array}
\right. .  \label{a45}
\end{equation}
Note that the structure of the Eq.s (\ref{a17}) and (\ref{a44})
are similar except the signs of the parameters. Accordingly, the
eigen solutions for $\lambda ,\nu ,E>0$ can be found from $\lambda
^{\prime },\nu ^{\prime },E^{\prime }<0$. \ Inserting the
relations (\ref{a45}) into the Eq. (\ref{a37}), one thus finds
that
\[
E=\lambda ^{2/(\nu +2)}\left( \frac{\hbar ^{2}}{2m}\right) ^{\nu /(\nu +2)}
\]
\begin{equation}
\cdot \left[ 2\nu \sqrt{\pi }\left( n+\frac{(q+\left| k+\mu _{0}\right| )}{2}%
+\frac{3}{4}\right) \cdot \frac{\Gamma \left( \frac{1}{\nu }+\frac{3}{2}%
\right) }{\Gamma \left( \frac{1}{\nu }\right) }\right] ^{2\nu /(\nu +2)}
\label{a46}
\end{equation}
with the ranges of the parameters $\lambda ,\nu
,E>0,\;n,q=0,1,2,\cdots ,$ and $-\infty <k<\infty $. As a
realization, the three-dimensional simple harmonic oscillator
moving in the presence of the A-B magnetic flux can be described
by the model with the parameters $\nu =2$ and $\lambda
=m\omega ^{2}/2$. Hence we can calculate the energy engenvalue from the Eq. (%
\ref{a46}) that gives us the following result
\begin{equation}
E_{n,q,k}=\left[ 2n+(q+\left| k+\mu _{0}\right| )+\frac{3}{2}\right] \hbar
\omega .  \label{a47}
\end{equation}
Another example is given by the model with an infinitely deep potential,
\begin{equation}
V(r)=\lambda r^{\nu }=\left\{
\begin{array}{l}
\infty ,
%\;\nu =\infty ,
\;\; {\rm for } \; r\geq a \\
0,\; \;\; {\rm for } \; r<a
\end{array}
\right. .  \label{a48}
\end{equation}
Similarly, Eq. (\ref{a46}) implies the following energy spectra
\begin{equation}
E_{n,q,k}=\frac{\hbar ^{2}\pi ^{2}}{2ma^{2}}\left[ n+\frac{q+\left| k+\mu
_{0}\right| }{2}+1\right] ^{2}.  \label{a49}
\end{equation}
Here we have replaced ($n+\gamma /2+3/4)$ with ($n+\gamma /2+1)$
according to the matching condition of the WKB approximation given
in the Appendix A. The analytic energy spectra of this system is
then given by the zeros of the modified Bessel function in the Eq.
(3.28) of the Ref. \cite{13}
\begin{equation}
I_{q+\left| k+\mu _{0}\right| +1/2}\left( \frac{\sqrt{-2mE}}{\hbar }a\right)
=0.  \label{a50}
\end{equation}

\input{epsf}
\begin{figure}[hbt]
\begin{center}
\epsfxsize=5in \epsffile{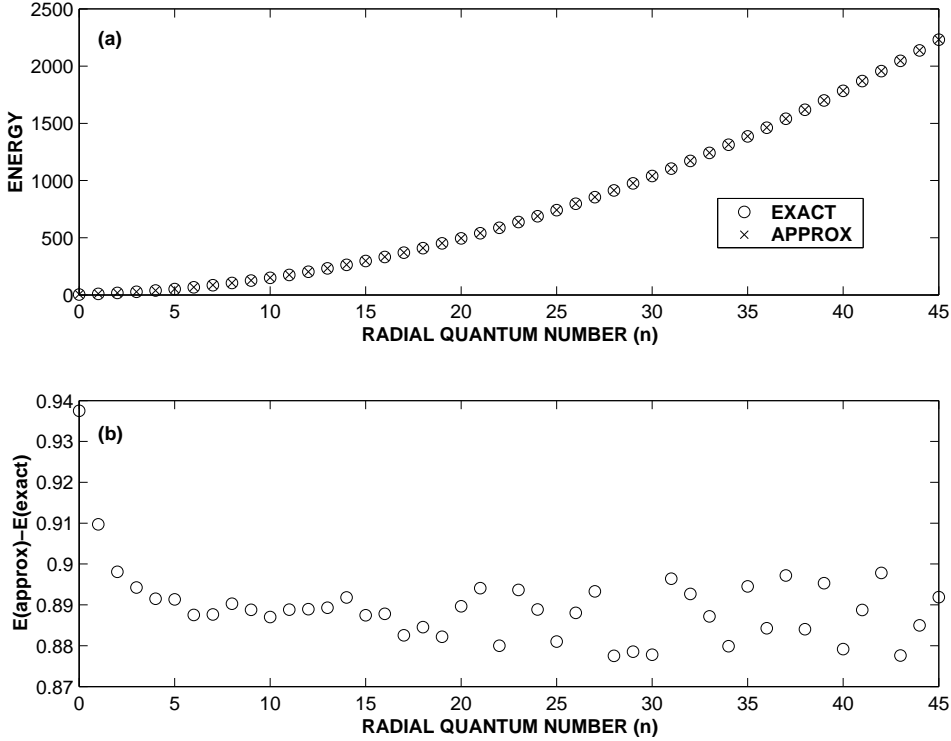}\vspace{5mm}
\caption{\label{figure1}\small Comparison of the exact and
approximate energy eigenvalue as a function of the radial quantum
number $n$. (a)The exact and approximate energy eigenvalue are
shown in Fig. 1(a). (b)Their difference is shown in Fig. 1(b).
Here we have set $q+|k+\mu_0|=2.5$. See Eq.s ~\ref{a49} and
\ref{a50} for details.The unit of the energy eigenvalue is set as
$\hbar ^{2}\pi ^{2}/2ma^{2}$. }
\end{center}
\end{figure}

The numerical analysis shown in Fig. 1 (a) (for $q+\left| k+\mu
_{0}\right| =2.5$) indicates that the result (\ref{a49}) is in
good agreement with the exact result (\ref{a50}). In addition,
Fig. 1 (b) exhibits the difference between the exact and
approximate results.
%%%%%%%%%%%%%%%%%%%%%%%%%%%%%%%%%%%%%%%%%%%%%%%%%%%%%%%%%%%%%%%
%\begin{figure}
%\includegraphics[width=11cm]{Fig1.eps}
%\caption{\label{fig:F1} (a) $(q+\left| k+\mu _{0}\right| )=2.5$
% (b) exhibits the difference between the exact and approximate
%results}
%\end{figure}
%%%%%%%%%%%%%%%%%%%%%%%%%%%%%%%%%%%%%%%%%%%%%%%%%%%%%%%%%%%%%%%

\section{The $\nu$-dependence of the distribution of the Energy Spectra }

Note that the Eq. (\ref{a46}) indicates that
\begin{equation}
E_{n,q,k}\propto \left( n+\frac{q+\left| k+\mu _{0}\right| }{2}+\frac{3}{4}%
\right) ^{2\nu /(\nu +2)}.  \label{b1}
\end{equation}
For example, for the model with an infinitely deep potential (i.e. $\nu
\rightarrow \infty ),$ one has
\begin{equation}
E_{n,q,k}\propto \left( n+\frac{q+\left| k+\mu _{0}\right| }{2}+1\right)
^{2}.  \label{b2}
\end{equation}
On the other hand, from the Eq. (\ref{a37}), one has (when $-2<\nu
<0)$
\begin{equation}
E_{n,q,k}\propto -\left[ \left( n+\frac{2\left( q+\left| k+\mu _{0}\right|
\right) +\nu +3}{2\nu +4}\right) \right] ^{2\nu /(\nu +2)}.  \label{b3}
\end{equation}
In addition, we can calculate their derivative with respect to $n$ and find
that
\begin{equation}
\frac{\partial E_{n,q,k}}{\partial n}>0  \label{b4}
\end{equation}
for all considered models. Thus one expects that the energy levels
$E_{n,q,k}$ will monotonically increase as $n$ monotonically
increases. The $q$ and $\left| k+\mu _{0}\right| $ dependence of
the energy eigenvalue $E_{n,q,k}$ can be found by the
Hellmann-Feynman formula (e.g. \cite{19})
\begin{equation}
\frac{\partial E_{n,q,k}}{\partial q}=\left\langle \Psi _{n,q,k}\left| \frac{%
\partial H}{\partial q}\right| \Psi _{n,q,k}\right\rangle ,  \label{b5}
\end{equation}
where the Hamiltonian is given by
\begin{equation}
H=-\frac{\hbar ^{2}}{2m}\frac{d^{2}}{dr^{2}}+\left( \lambda r^{\nu }+\frac{%
\hbar ^{2}}{2m}\frac{(q+\left| k+\mu _{0}\right| )(q+\left| k+\mu
_{0}\right| +1)}{r^{2}}\right) .  \label{b6}
\end{equation}
Thus, we can derive the following results:
\begin{equation}
\frac{\partial E_{n,q,k}}{\partial q}=\left\langle \Psi _{n,q,k}\left| \frac{%
\left[ 2(q+\left| k+\mu _{0}\right| )+1\right] \hbar ^{2}}{2mr^{2}}\right|
\Psi _{n,q,k}\right\rangle >0,  \label{b7}
\end{equation}
\begin{equation}
\frac{\partial E_{n,q,k}}{\partial \left| k+\mu _{0}\right|
}=\left\langle \Psi _{n,q,k}\left| \frac{\left[ 2(q+\left| k+\mu
_{0}\right| )+1\right] \hbar ^{2}}{2mr^{2}}\right| \Psi
_{n,q,k}\right\rangle >0.  \label{b8}
\end{equation}
This means that the energy spectra $E_{n,q,k}$ will monotonically
increase as any one of the quantum numbers in the set ($n,q,k)$
monotonically increases. Therefore the ground state will be given
by $n=q=k=0$. The details can be obtained by analyzing the
tendency of $E_{n,q,k}$ with respect to the change of the
parameter $\nu $.

\input{epsf}
\begin{figure}[hbt]
\begin{center}
\epsfxsize=4.5in \epsffile{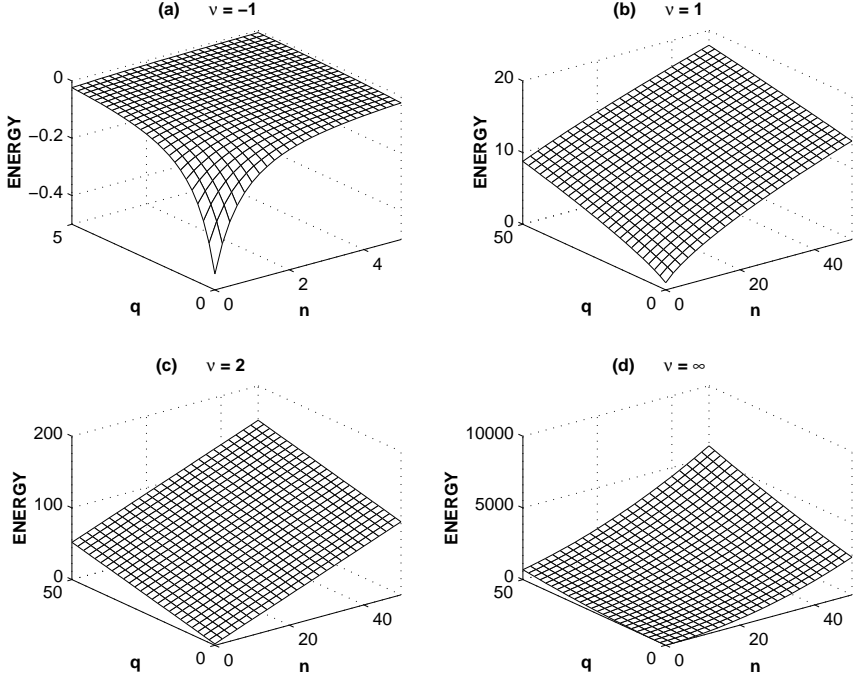}\vspace{5mm}
\caption{\label{figure2}\small Energy as a function of $q$ and $n$
for four different $\nu$'s are shown in Fig. 2. Here we have
chosen $|k+\mu_0|=0.5$. The unit of the energy eigenvalue is set
as $mc^{2}\alpha ^{2}/2$, $\left( 9\pi ^{2}\lambda ^{2}\hbar
^{2}/8m\right) ^{1/3}$, $\hbar \omega $, and  $\hbar ^{2}\pi
^{2}/2ma^{2}$ for Fig.2(a), Fig.2(b), Fig.2(c), and Fig.2(d)
respectively.}
\end{center}
\end{figure}

\subsection{Distribution Tendency of The Energy Spectra for $\protect\nu =-1$%
}

The energy spectra for a charged particle moving in the Coulomb potential
and an A-B flux is given by the Eq. (\ref{a38}). Its first and second order
derivatives with respect to the parameters $(n,q,\left| k+\mu _{0}\right| )$
are
\begin{equation}
\left\{
\begin{array}{l}
\frac{\partial E_{n,q,k}}{\partial n}=mc^{2}\alpha ^{2}\frac{1}{(n+q+\left|
k+\mu _{0}\right| +1)^{3}}>0 \\
\frac{\partial ^{2}E_{n,q,k}}{\partial n^{2}}=mc^{2}\alpha ^{2}\frac{-3}{%
(n+q+\left| k+\mu _{0}\right| +1)^{4}}<0
\end{array}
\right. ,  \label{b12}
\end{equation}
\begin{equation}
\left\{
\begin{array}{l}
\frac{\partial E_{n,q,k}}{\partial q}=mc^{2}\alpha ^{2}\frac{1}{(n+q+\left|
k+\mu _{0}\right| +1)^{3}}>0 \\
\frac{\partial ^{2}E_{n,q,k}}{\partial q^{2}}=mc^{2}\alpha ^{2}\frac{-3}{%
(n+q+\left| k+\mu _{0}\right| +1)^{4}}<0
\end{array}
\right. ,  \label{b13}
\end{equation}
\begin{equation}
\left\{
\begin{array}{l}
\frac{\partial E_{n,q,k}}{\partial \left| k+\mu _{0}\right| }=mc^{2}\alpha
^{2}\frac{1}{(n+q+\left| k+\mu _{0}\right| +1)^{3}}>0 \\
\frac{\partial ^{2}E_{n,q,k}}{\partial \left| k+\mu _{0}\right| ^{2}}%
=mc^{2}\alpha ^{2}\frac{-3}{(n+q+\left| k+\mu _{0}\right| +1)^{4}}<0
\end{array}
\right. .  \label{b14}
\end{equation}
Consequently, $E_{n,q,k}$ tends to increase and saturate gradually
as anyone of the parameters in the set ($n,q,\left| k+\mu
_{0}\right| )$ increases. It implies the bending curve as shown in
Fig.2 (a). The unit of the energy eigenvalue in Fig. 2 (a) is
chosen as $mc^{2}\alpha ^{2}/2$.

\subsection{Distribution Tendency of the Energy Spectra for $\protect\nu =1$}

The energy levels for the model with $\nu =1$ are given by the Eq. (\ref{a46}%
)
\begin{equation}
E_{n,q,k}=\left( \frac{\lambda ^{2}\hbar ^{2}}{2m}\right) ^{1/3}\left[ \frac{%
3\pi }{2}\left( n+\frac{(q+\left| k+\mu _{0}\right| )}{2}+\frac{3}{4}\right) %
\right] ^{2/3}.  \label{b19}
\end{equation}
Their derivatives with respect to the parameters ($n,q,\left|
k+\mu _{0}\right| )$ yield
\begin{equation}
\left\{
\begin{array}{l}
\frac{\partial E_{n,q,k}}{\partial n}=\left( \frac{\lambda ^{2}\hbar ^{2}}{2m%
}\right) ^{1/3}\pi \left[ \frac{3\pi }{2}\left( n+\frac{(q+\left| k+\mu
_{0}\right| )}{2}+\frac{3}{4}\right) \right] ^{-1/3}>0,\; \\
\frac{\partial ^{2}E_{n,q,k}}{\partial n^{2}}=-\left( \frac{\lambda
^{2}\hbar ^{2}}{2m}\right) ^{1/3}\left( \frac{\pi ^{2}}{2}\right) \left[
\frac{3\pi }{2}\left( n+\frac{(q+\left| k+\mu _{0}\right| )}{2}+\frac{3}{4}%
\right) \right] ^{-4/3}<0
\end{array}
\right. ,  \label{b20}
\end{equation}
\begin{equation}
\left\{
\begin{array}{l}
\frac{\partial E_{n,q,k}}{\partial q}=\left( \frac{\lambda ^{2}\hbar ^{2}}{2m%
}\right) ^{1/3}\frac{\pi }{2}\left[ \frac{3\pi }{2}\left( n+\frac{(q+\left|
k+\mu _{0}\right| )}{2}+\frac{3}{4}\right) \right] ^{-1/3}>0,\; \\
\frac{\partial ^{2}E_{n,q,k}}{\partial q^{2}}=-\left( \frac{\lambda
^{2}\hbar ^{2}}{2m}\right) ^{1/3}\left( \frac{\pi ^{2}}{8}\right) \left[
\frac{3\pi }{2}\left( n+\frac{(q+\left| k+\mu _{0}\right| )}{2}+\frac{3}{4}%
\right) \right] ^{-4/3}<0
\end{array}
\right. ,  \label{b21}
\end{equation}
and
\begin{equation}
\left\{
\begin{array}{l}
\frac{\partial E_{n,q,k}}{\partial \left| k+\mu _{0}\right| }=\left( \frac{%
\lambda ^{2}\hbar ^{2}}{2m}\right) ^{1/3}\left( \frac{\pi }{2}\right) \left[
\frac{3\pi }{2}\left( n+\frac{(q+\left| k+\mu _{0}\right| )}{2}+\frac{3}{4}%
\right) \right] ^{-1/3}>0,\; \\
\frac{\partial ^{2}E_{n,q,k}}{\partial \left| k+\mu _{0}\right| ^{2}}%
=-\left( \frac{\lambda ^{2}\hbar ^{2}}{2m}\right) ^{1/3}\left( \frac{\pi ^{2}%
}{8}\right) \left[ \frac{3\pi }{2}\left( n+\frac{(q+\left| k+\mu _{0}\right|
)}{2}+\frac{3}{4}\right) \right] ^{-4/3}<0
\end{array}
\right. .  \label{b22}
\end{equation}
It is obvious that $E_{n,q,k}$ will monotonically increase when
the value of any parameter of the set $(n,q,\left| k+\mu
_{0}\right| )$ increases as shown in Fig.2 (b). Note that the
slope is much more smooth than the model with $\nu =-1$. The unit
of energy in Fig.2(b) is chosen as $\left( 9\pi ^{2}\lambda
^{2}\hbar ^{2}/8m\right) ^{1/3}$.

\subsection{Distribution Tendency of the Energy Spectra for $\protect\nu =2$}

The energy spectra for a charged particle moving in the three-dimensional
harmonic potential and an A-B flux is given by the Eq. (\ref{a47}). Its
first and second order derivatives with respect to the set of parameters $%
(n,q,\left| k+\mu _{0}\right| )$ read
\begin{equation}
\frac{\partial E_{n,q,k}}{\partial n}=2\hbar \omega \;({\rm const.}),\;\frac{%
\partial ^{2}E_{n,q,k}}{\partial n^{2}}=0,  \label{b91}
\end{equation}
\begin{equation}
\frac{\partial E_{n,q,k}}{\partial q}=\hbar \omega \;({\rm const.}),\;\frac{%
\partial ^{2}E_{n,q,k}}{\partial q^{2}}=0,  \label{b10}
\end{equation}
and
\begin{equation}
\frac{\partial E_{n,q,k}}{\partial \left| k+\mu _{0}\right| }=\hbar \omega
\;({\rm const.}),\;\frac{\partial ^{2}E_{n,q,k}}{\partial \left| k+\mu
_{0}\right| ^{2}}=0.  \label{b11}
\end{equation}
This means that $E_{n,q,k}$ will linearly increase as any one of
the parameter in the set $(n,q,\left| k+\mu _{0}\right| )$
increases. The details is shown in Fig.2 (c) with the unit of
energy given by $\hbar \omega $.

\subsection{Distribution Tendency of the Energy Spectra for $\protect\nu %
=\infty $}

According to the Eq. (\ref{a49}), we obtain the first and second
order derivatives
with respect to $E_{n,q,k}$%
\begin{equation}
\frac{\partial E_{n,q,k}}{\partial n}=\frac{\pi ^{2}\hbar ^{2}}{ma^{2}}\left[
n+\frac{(q+\left| k+\mu _{0}\right| )}{2}+1\right] >0,\;\frac{\partial
^{2}E_{n,q,k}}{\partial n^{2}}=\frac{\pi ^{2}\hbar ^{2}}{ma^{2}}>0,
\label{b15}
\end{equation}
\begin{equation}
\frac{\partial E_{n,q,k}}{\partial q}=\frac{\pi ^{2}\hbar ^{2}}{2ma^{2}}%
\left[ n+\frac{(q+\left| k+\mu _{0}\right| )}{2}+1\right] >0,\;\frac{%
\partial ^{2}E_{n,q,k}}{\partial q^{2}}=\frac{\pi ^{2}\hbar ^{2}}{4ma^{2}}>0,
\label{b16}
\end{equation}
and
\begin{equation}
\frac{\partial E_{n,q,k}}{\partial \left| k+\mu _{0}\right| }=\frac{\pi
^{2}\hbar ^{2}}{2ma^{2}}\left[ n+\frac{(q+\left| k+\mu _{0}\right| )}{2}+1%
\right] >0,\;\frac{\partial ^{2}E_{n,q,k}}{\partial \left| k+\mu _{0}\right|
^{2}}=\frac{\pi ^{2}\hbar ^{2}}{4ma^{2}}>0,  \label{b18}
\end{equation}
for the model with $\nu \to \infty$. Note that $E_{n,q,k}$ will
monotonically increase when any one of the parameters in the set $%
(n,q,\left| k+\mu _{0}\right| )$ increases. The rate of increase
is, however, faster than the model $\nu =2$ since the curve climbs
up as shown in Fig.2 (d) with the unit chosen as $\hbar ^{2}\pi
^{2}/2ma^{2}$.

In summary, all these results imply the following rules for a
charged particle moving in the spherically symmetric potential
$V(r)=\lambda r^{\nu } $ ($-2<\nu <\infty $) \cite {11a} and an
A-B magnetic flux:

(a) The energy spectra of the bound states depend on the quantum
number ($n,q,k)$ and monotonically increase as any one of the
quantum numbers increases.

(b) when $\nu =2$, the energy spectra $E_{n,q,k}$ depend linearly
on any parameter in the set ($n,q,k)$; when $\nu >2$, the energy
curve bends up
as any one of the quantum numbers ($n,q,k)$ increases. On the other hand, when $%
\nu <2$, the curve bends down as any one of the quantum numbers
increases.

(c) when $\nu =2$, we have $\partial E/\partial n:\partial
E/\partial q=2:1,$ $\partial E/\partial n:\partial E/\partial
\left| k+\mu _{0}\right| =2:1,$ and $\partial E/\partial
q:\partial E/\partial \left| k+\mu _{0}\right| =1:1$ which are
related to the closeness of the classical orbits and whether the
model is exactly solvable or not. For the case with positive power
of $\nu $ ,
\[
E\sim \left[ \left( n+\frac{(q+\left| k+\mu _{0}\right|
)}{2}+\frac{3}{4} \right) \right] ^{2\nu /(\nu +2)} .
\]
Although we still have the same ratio of derivatives, the above
relation does not hold for the exact solution.

(d) when $\nu =-1,$ its energy spectra have the properties,
$\partial E/\partial n:\partial E/\partial q=1:1,$ $\partial
E/\partial n:\partial E/\partial \left| k+\mu _{0}\right| =1:1,$
and $\partial E/\partial q:\partial E/\partial \left| k+\mu
_{0}\right| =1:1$. They are also related to the closeness of the
classical orbits. For the models with negative power of $\nu
,(-2<\nu <0)$, the WKB approximation given by the Eq. (\ref {a37})
implies that
\[
E\sim \left[ \left( n+\frac{2(q+\left| k+\mu _{0}\right| )+\nu +3}{ 2\nu +4}%
\right) \right] ^{2\nu /(\nu +2)} .
\]
This hence implies that $\partial E/\partial n:\partial E/\partial
q=\nu +2,$ $\partial E/\partial n:\partial E/\partial \left| k+\mu
_{0}\right| =\nu +2,$ and $\partial E/\partial q:\partial
E/\partial \left| k+\mu _{0}\right| =\nu +2$ are all equal. This
relation does not hold for the exact result for the same reason.

(e) The increase of intensity of the magnetic flux will change the
slope of the energy distribution in both the models with $-2<\nu
<0$ and the models with $0<\nu <\infty $. More explicitly, when
$\nu <2$, increasing the flux will depress the slope; whereas when
$\nu >2,$ increasing the flux will lead to the increase of the
slope. In addition, the model with $\nu =2$ is marginal in the
sense that the slope of the energy distribution will not be
affected by the change of the flux. For details, see the
difference shown in the Fig.2 and Fig.3. Note that $|k+ \mu_0|$ is
set as $0.5$ and $12$ in Fig.2 and Fig.3 respectively.

\input{epsf}
\begin{figure}[hbt]
\begin{center}
\epsfxsize=5in \epsffile{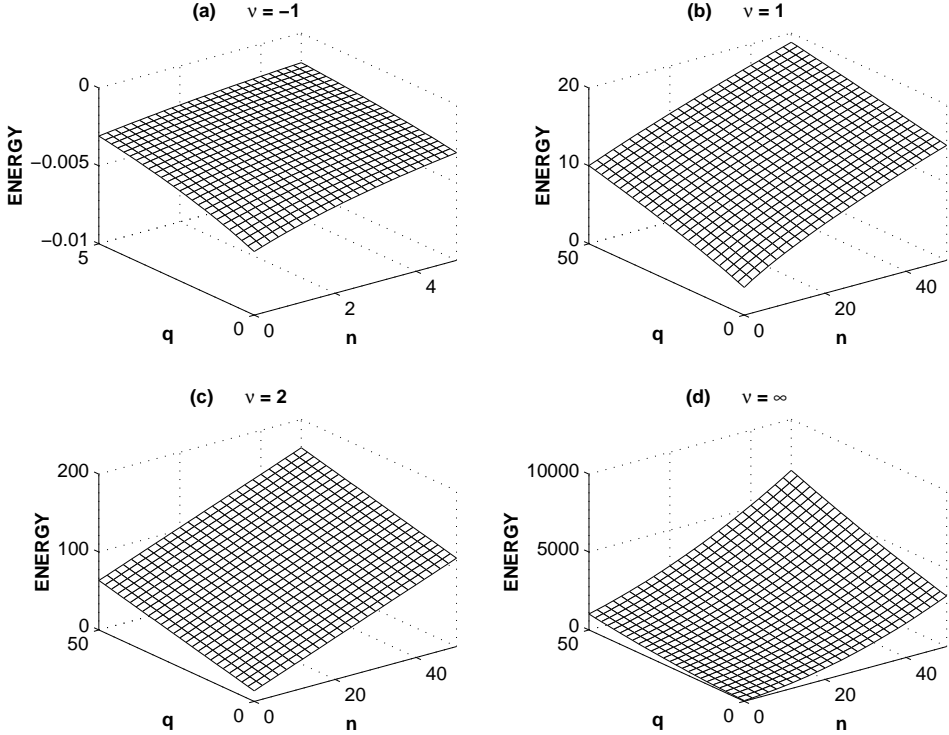}\vspace{5mm}
\caption{\label{figure3}\small Energy as a function of $q$ and $n$
for four different $\nu$'s. Here we choose $|k+\mu_0|=12$.The unit
of the energy eigenvalue is set as $mc^{2}\alpha ^{2}/2$, $\left(
9\pi ^{2}\lambda ^{2}\hbar ^{2}/8m\right) ^{1/3}$, $\hbar \omega
$, and  $\hbar ^{2}\pi ^{2}/2ma^{2}$ for Fig.3(a), Fig.3(b),
Fig.3(c), and Fig.3(d) respectively. }
\end{center}
\end{figure}

\section{Conclusion}

The semiclassical quantization rule is presented for a charged
particle moving in a system with a general central force described
by the potential $V(r)=\lambda r^{\nu \text{ }}$, with $-2<\nu
<\infty $, and an A-B magnetic flux. The formulae obtained in this
paper are in good agreement with the energy levels with all known
exactly solvable models with some specific values of $\nu $.
Furthermore, we have presented numerical results for $\nu =\infty
$ which is also in good agreement with the exact result.
Therefore, one expects that the semiclassical quantization rules
will also be in good agreement with the models prescribed by a
large ranges of $\nu $ even the results shown in this paper are
more reliable for the case with large principle quantum number
$n$.
\section{Appendix A}

The WKB wave function for a charged particle moving in a smooth
potential well near the neighborhood $x\sim a$ $(x>a)$, where
$x=a,b$ are the intersection points of the horizonal line $y=E$
and the curve $y=V(x)$ as shown in Fig. 4(a), can be expressed in
terms of the classical momentum $p$ as (see, for example, Ref.
\cite{19} for details)
\begin{equation}
\Psi (x)=\frac{C}{\sqrt{p}}\sin \left[ \frac{1}{\hbar }\int_{a}^{x}pdx+\frac{%
\pi }{4}\right] \equiv \frac{C}{\sqrt{p}}\sin \alpha (x),
\label{c1}
\end{equation}
were $C$ is constant. Analogously, near the neighborhood $x\sim b$ $(x<b)$
we have
\begin{equation}
\Psi (x)=\frac{C^{\prime }}{\sqrt{p}}\sin \left[ \frac{1}{\hbar }%
\int_{x}^{b}pdx+\frac{\pi }{4}\right] \equiv \frac{C^{\prime }}{\sqrt{p}}%
\sin \beta (x).  \label{c2}
\end{equation}
These two wave functions must be consistent. This means that near the
neighborhoods $a,b$ of $x$
\begin{equation}
\alpha (x)+\beta (x)=\frac{1}{\hbar }\int_{a}^{b}pdx+\frac{\pi
}{2}=(n+1)\pi ,\;n=0,1,2,3\cdots .  \label{c3}
\end{equation}
Or equivalently,
\begin{equation}
\oint pdx=(n+\frac{1}{2})h,\;n=0,1,2,3\cdots .  \label{c4}
\end{equation}
For the half infinite potential well as shown in Fig. 4(b), one has
\begin{equation}
\oint pdx=(n+\frac{3}{4})h,\;n=0,1,2,3\cdots .  \label{c5}
\end{equation}
Analogously, the matching rule of the wave functions gives the
quantization rule for the system with an infinitely deep
square-well potential as illustrated in Fig. 4(c). Indeed, one has
\begin{equation}
\oint pdx=(n+1)h,\;n=0,1,2,3\cdots .  \label{c6}
\end{equation}
The argument leading to the same result for a more general
condition beyond the above examples can be found with the help of
the Maslov index shown in the Ref. \cite{20}.

\input{epsf}
\begin{figure}[hbt]
\begin{center}
\epsfxsize=5in \epsffile{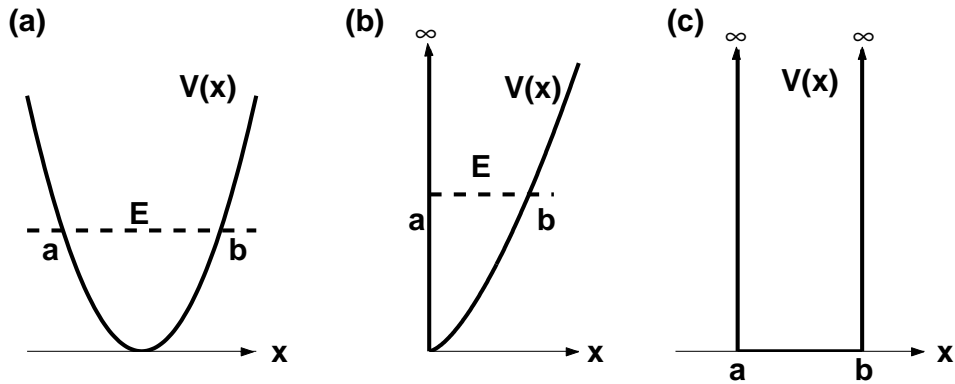}\vspace{5mm}
\caption{\label{figureA}\small
 WKB wave function matching boundary
conditions for three cases of potentials.}.
\end{center}
\end{figure}

%%%%%%%%%%%%%%%%%%%%%%%%%%%%%%%%%%%%%%%%%%%%%%%%%%%%%%%%%%%%%%%%%%%%%%
\section*{Acknowledgments}
%%%%%%%%%%%%%%%%%%%%%%%%%%%%%%%%%%%%%%%%%%%%%%%%%%%%%%%%%%%%%%%%%%%%%%
This work is supported in part by the National Science Council
under the grant numbers NSC90-2112-M009-021.


\begin{references}
\bibitem{1}  M.G. Alford and F. Wilczek, Phys. Rev. Lett. {\bf 62}, 1071
(1989).

\bibitem{2}  D. Deser, R. Jackiw and G.'tHooft, Ann. Phys. {\bf 152}, 220
(1984); S. Deser and R. Jackiw, Commun. Math. Phys. {\bf 118}, 495 (1988);
P. de Sousa Gerbert and R. Jackiw, Commun. Math. Phys. {\bf 124}, 229 (1989).

\bibitem{3}  F. Wilczek, Phys. Rev. Lett. 49, 957 (1982); Y.H. Chen, F.
Welczek, E. Witten and B.I. Halperin, Int. J. Mod. Phys. {\bf B3}, 1001
(1989).

\bibitem{4}  Z.F. Ezawa, M. Hotta, and A. Iwazaki, Phys. Rev. {\bf D 44},
3906 (1991).

\bibitem{5}  C.A. Trugenberger, Phys. Rev. {\bf D 45}, 3807 (1992).

\bibitem{6}  R.B. Laughlin, Phys. Rev. {\bf B 23}, 3383 (1983); F.D.M.
Haldane, Phys. Rev. Lett. {\bf 51}, 605 (1983); B.I. Halperin, Phys. Rev.
Lett. {\bf 52}, 1583 (1984).

\bibitem{7}  Z.F. Ezawa and A. Iwazaki, Phys. Rev. {\bf B 43}, 2637 (1991).

\bibitem{8}  R.B. Laughlin, Phys. Rev. Lett. {\bf 60}, 1057 (1988); A.
Fetter, C. Hanna, and R.B. Laughlin, Phys. Rev. B {\bf 39}, 9679 (1989).

\bibitem{9}  I.V. Barashenkov and A.O. Harin, Phys. Rev. Lett. {\bf 72},
1575 (1994); Phys. Rev. {\bf D 52}, 2471 (1995).

\bibitem{10}  A. Guha and S. Mukherjee, J. Math. Phys. {\bf 28}, 840 (1987);
M. Kibler and T. Negadi, Phys. Lett. {\bf A 124}, 42 (1987); G.E.
Draganascu, C. Campigotto, and M. Kibler, Phys. lett. {\bf A 170}, 339
(1992); V.M. villalba, Phys. Lett. {\bf A 193}, 218 (1994); L. Chetonani, L.
Guechi, and T.F. Hamman, J. Math. Phys. {\bf 30}, 655 (1989).

\bibitem{11}  D.H. Lin, J. Phys. {\bf A}, 4785 (1998); J. Math. Phys. {\bf 40%
}, 1264 (1999).

\bibitem{12}  Q.G. Lin, Phys. Rev. {\bf A 59}, 3228 (1999).

\bibitem{Gu}  M.C. Gutzwiller: {\it Chaos in Classical and Quantum Mechanics}
(Springer Verlag, New York, 1990).

\bibitem{St}  V.M. Strutinsky, Nukleonika (Poland): {\bf 20}, 679 (1975);
V.M. Strutinsky and A.G. Manger, Sov. J. Part. Nucl. {\bf 7}, 138 (1976).

\bibitem{Ma}  M. Brack and R.K. Bhaduri: {\it Semiclassical Physics}
(Addison-Wesley, New York, 1997).

\bibitem{13}  D.S. Chuu and D.H. Lin, J. Phys. {\bf A}, 34 (2001).

\bibitem{14}  C.N. Yang, Phys. Rev. Lett, {\bf 33}, 445 (1974).

\bibitem{15}  T.T. Wu and C.N. Yang, Phys. Rev, {\bf D12}, 3845 (1975).

\bibitem{16}  D.H. Lin, J. Math. Phys. {\bf 41,} 2723 (2000).

\bibitem{17}  D.H. Lin, Ann. Phys. {\bf 290}, 1 (2001).

\bibitem{17a}  H. Kleinert, {\it Path Integrals in Quantum Mechanics,
Statistics and Polymer Physics}, World Scientific, Singapore, 1995.

\bibitem{11a}  J.Y. Zeng, {\it Problem in Quantum Mechanics} (Science Press,
Beijing, 1988).

\bibitem{18}  W. Magnus, F. Oberhettinger and R.P. Soni, {\it Formulas and
Theorems of the Special Function of Mathematical Physics} ( Springer,
Berlin, 1966).

\bibitem{19}  J.Y. Zeng, {\it Quantum Mechanics }(Science Press, Beijing,
1999).

\bibitem{20}  H. Kleinert and D.H. Lin, quant-ph/9807068; D.H. Lin,
quant-ph/9901049.
\end{references}
\end{document}